# Integrated simulation of cavity design and radiation transport codes (ACE3P + Geant4)*

*Lixin Ge, Zenghai Li, Cho-Kuen Ng, Liling Xiao*
*SLAC, Stanford, CA 94025, USA*

*Hiroyasu Ego, Yoshinori Enomoto, Hiroshi Iwase, Yu Morikawa, Takashi Yoshimoto*
*KEK , 1-1 Oho, Tsukuba, Ibaraki, Japan*

## ABSTRACT

A simulation workflow has been developed to study dark current (DC) radiation effects using ACE3P and Geant4. The integrated workflow interfaces particle data transfer and geometry between the electromagnetic (EM) cavity simulation code ACE3P and the radiation code Geant4, targeting large-scale problems using high-performance computing. The process begins by calculating the operating mode in the vacuum region of an accelerator structure and tracking field-emitted electrons influenced by the EM fields of the mode calculated by ACE3P. It then transfers particle data at the vacuum-wall interface for subsequent radiation calculations within the wall enclosure materials through Geant4 calculation. The whole integrated simulation workflow will be demonstrated through large-scale dark current radiation calculations for the KEK 56-cell traveling-wave structure, and the efficiency of performing these simulations on the NERSC supercomputer Perlmutter will be presented.

## 1. INTRODUCTION

Accelerator cavities operating at high gradients are vulnerable to damage caused by high-energy electrons striking the cavity wall. These electrons are believed to originate from specific locations on the cavity surface due to field emission. Once emitted, they are accelerated by the cavity's RF field before colliding with the wall. Upon impact, these electrons interact with the cavity wall, which is composed of certain materials, generating electromagnetic radiation. This radiation can degrade the performance of an accelerator.

Appears in the proceedings of the 14th International Computational Accelerator Physics Conference (ICAP'24), 2-5 October 2024, Germany.
* Supported by HEP US - Japan Science and Technology Cooperation Program (2022-2025).

The current approach to radiation calculations involves separate simulations and requires expertise from multiple physics domains. Additionally, high computational performance is essential for large accelerator systems. Therefore, it is crucial to develop an easy-to-use, streamlined tool that enables the accelerator community to better utilize available resources.

Supported by the HEP US-Japan Science and Technology Cooperation Program, we integrate ACE3P [1-5], for modelling accelerator structures, and Geant4 [6-8], for simulating particle interactions with matter, into a unified simulation workflow. This streamlined approach eliminates the need for individual scientists to perform separate tasks and coordinate extensively with one another.

This paper builds upon and updates the work presented in [9].

## 2. ACE3P-GEANT4 INTEGRATION

### 2.1 ACE3P

ACE3P is a comprehensive suite of conformal, high-order, C++/MPI-based parallel finite-element (FE) multiphysics codes, offering electromagnetic (EM), thermal, and mechanical simulation capabilities. Developed at SLAC over nearly two decades with the support of DOE Computational Grand Challenge and SciDAC [10] initiatives, ACE3P encompasses a range of specialized modules. These include frequency domain solvers like Omega3P for eigenmode analysis and S3P for S-parameter calculations; time-domain solvers such as T3P for wakefield and transient simulations; and particle tracking modules Track3P for multipacting and dark current simulations. Additional capabilities include Pic3P for RF gun and space charge effect studies, TEM3P for coupled EM, thermal, and mechanical analyses, and the recently developed static particle-in-cell solver, Gun3P, tailored for DC gun and space charge effect simulations.

ACE3P utilizes curved high-order finite elements for high-fidelity modelling and is implemented on massively parallel computers, such as those at NERSC [11], leveraging thousands or more cores to handle larger problem sizes and increase computational speed. By taking advantage of state-of-the-art computing architectures, the performance of ACE3P has significantly improved. Fig. 1 compares the performance of the Cori and Perlmutter supercomputers at NERSC for running Omega3P with 12 million degrees of freedom (DoFs). On Cori, the simulation takes over 40 minutes using 40 nodes, whereas on Perlmutter, it completes in under 20 minutes with only 3 nodes.

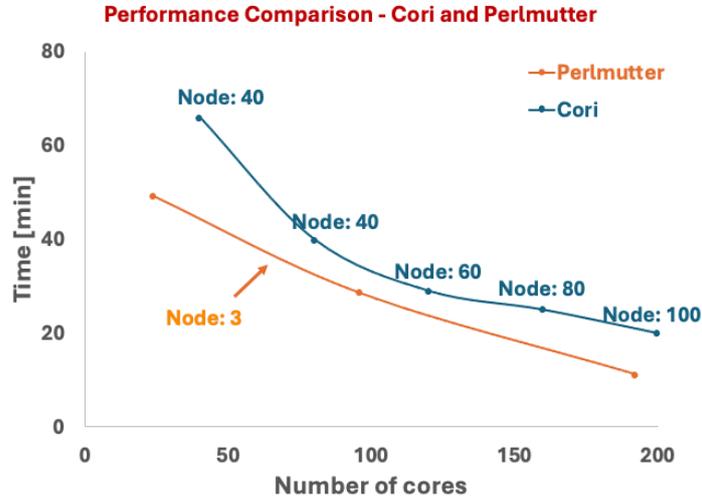

Figure 1. Performance comparison between Cori and Perlmutter for Omega3P with 12M DoFs.

ACE3P has been well accepted by the accelerator community as a benchmark and guidance of structure optimizations from large-scale simulations. In this integrated tool, Omega3P/S3P will be used for EM field calculation, and Track3P for DC simulation.

**2.2 Geant4**

Geant4 is a versatile software toolkit designed for simulating the interaction of elementary particles with matter. It is developed and maintained by the Geant4 Collaboration and serves as a critical tool for numerous high-energy physics experiments. Today, all LHC experiments at CERN, as well as many current and upcoming experiments at DOE laboratories, rely on Geant4 for accurate modeling. Recently, a Geant4-based positron beam source package (GPos) was developed at LBNL [12]. This publicly available C++ code is user-friendly and incorporates hybrid MPI support, openPMD [13], and parallel I/O capabilities, making it highly suitable for modeling the interactions between relativistic particle beams and solid targets.

**2.3 Simulation Workflow**

Fig. 2 illustrates a schematic of the simulation workflow for the integrated software. The process begins with constructing geometrical models, defining two separate computational domains for ACE3P and Geant4 simulations. An integrated code driver assigns the problem type, determining which code initiates the simulation: ACE3P for dark current analysis or Geant4 for radiation and positron source simulations. Particles that interact with the geometrical interface between the two domains are collected and transferred to the corresponding code for further physics simulation. The workflow concludes with the output of particle data and radiation distributions, which are stored in files for visualization and postprocessing.

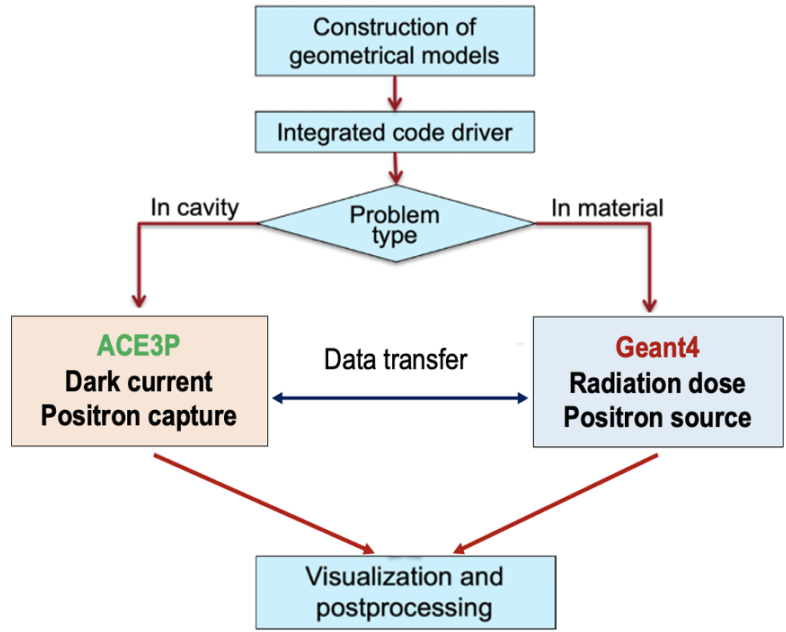

Figure 2. Workflow for integration of ACE3P and Geant4 simulation.

### 2.4 Particle Data I/O

A C++ API for openPMD [14] has been developed in ACE3P to convert unstructured electromagnetic field data in NetCDF format, generated from finite element discretization, into structured data suitable for Cartesian grid simulations used by other codes, such as the beam dynamics code IMPACT [15-17] and Geant4. The openPMD-API supports parallel execution using multiple compute cores with MPI, enabling faster data output.

For particle data transfer from ACE3P to Geant4, an intermediary particle data reader/writer has been implemented to handle 6D phase space data (position and momentum, x and p) and timestamps (t) in ASCII format. Plans are underway to expand this particle data transfer capability by incorporating openPMD-API readers/writers to further improve I/O efficiency.

### 2.5 CAD Model

Both ACE3P and Geant4 begin simulations from a CAD model of a geometric domain. The ACE3P modules—Omega3P, S3P, and Track3P—discretize the CAD model into a finite element mesh composed of curved finite elements. They also provide information about which finite element entities correspond to specific boundaries of the CAD model. ACE3P handles CAD geometry through the third-party mesh generator Cubit [19], which reads CAD models and generates finite element meshes for ACE3P simulations.

In contrast, Geant4 uses a faceted representation of the CAD model boundary. For model import into Geant4, a conversion tool is required to transform CAD models into the GDML format supported by Geant4. Several conversion tools have been developed for this purpose [20-23]. In this integration, Cubit is used to convert CAD models into the STL format, which Geant4 can recognize via CADMesh [24], a direct CAD model import interface for Geant4.

## 3. APPLICATION ON LARGE SCALE S-BAND STRUCTURE

KEK conducted DC simulations on a smaller 7-cell model, rather than the full 56-cell structure, using the commercial EM code CST due to limitations in computing power and memory. To validate ACE3P simulations, we benchmarked its DC simulation results against CST using the same 7-cell structure [9]. Additionally, KEK carried out dark current and radiation intensity experiments on a 56-cell S-band accelerating structure [25]. Fig. 3 shows the bench test of the S-band structure conducted at KEK. The complete integrated simulation workflow developed by SLAC will be demonstrated through large-scale dark current radiation calculations for the KEK 56-cell traveling-wave structure, with an emphasis on highlighting the computational efficiency achieved on the NERSC supercomputer Perlmutter.

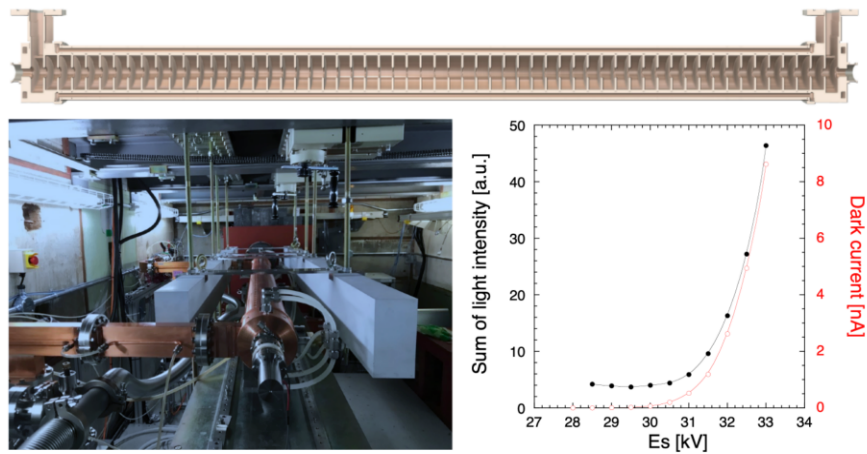

Figure 3. Bench test of an S-band structure at KEK.

**3.1 Model, mesh and EM field**

Based on the S-band 56-cell model provided by KEK (Fig. 4), a series of meshes with curved tetrahedral elements are generated for the full 56-cell vacuum region using Cubit. A finer mesh resolution is applied around each cell's iris region (Fig. 5). S3P, the S-parameter module in ACE3P, is employed to compute the operating mode at frequency of 2.856 GHz. Additionally, a mesh convergence study is conducted on the NERSC Perlmutter supercomputer (Fig. 6). For a problem with 3.4M curved tetrahedral elements, it took 3.4 minutes to get $2^{nd}$ order EM field by using 4 CPU nodes, 64 cores/node on NERSC Perlmutter.

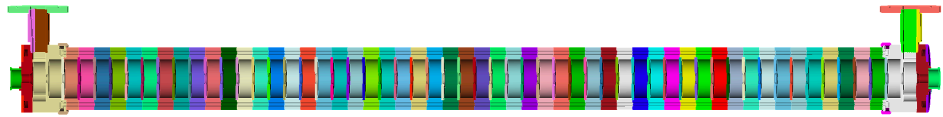

Figure 4. S-band 56-cell CAD model provided by KEK.

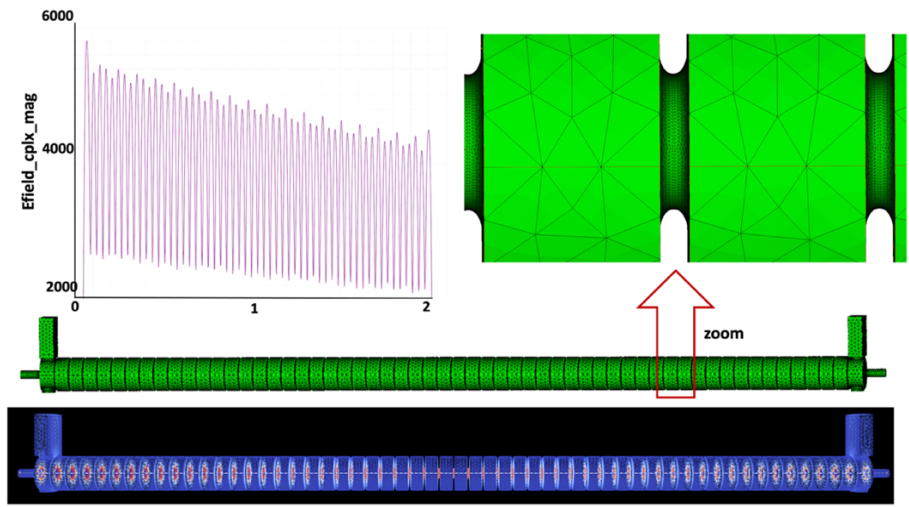

Figure 5. Mesh and complex E field magnitude profile.

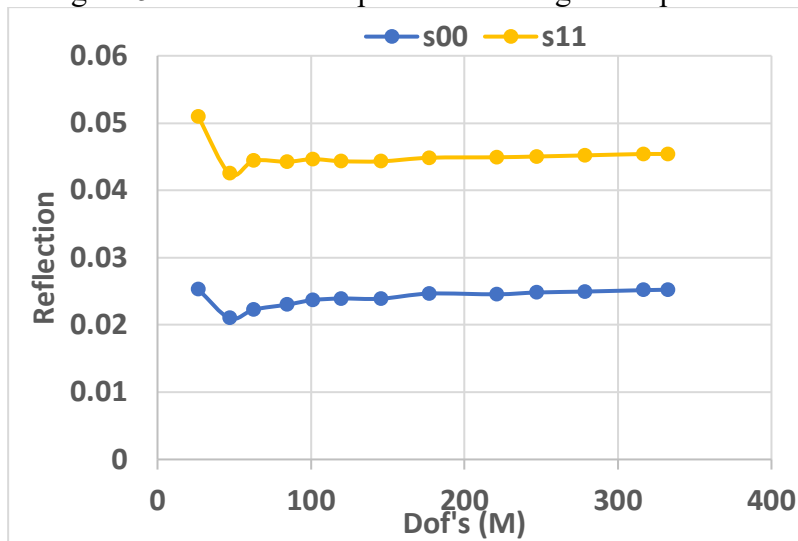

Figure 6. Mesh convergence study (s00 and s11 are scattering parameters).

## 3.2 Dark current simulation

Track3P, a particle tracking code in ACE3P, is used for dark current simulation. Fig. 7 shows dark current simulation workflow. Electrons are emitted from the cavity surface according to the Fowler-Nordheim formula

$$J(r,t) = 1.54 \times 10^{\left(-6+\frac{4.52}{\sqrt{\phi}}\right)} \frac{(\beta E)^2}{\phi} e^{\left(\frac{-6.53 \times 10^9 \phi^{1.5}}{\beta E}\right)}.$$

The high-fidelity geometry representation built in the finite-element method allows for realistic modeling of particle emission on the cavity wall. These electrons contribute to the dark current, and their movements in the cavity are governed by the rf fields, following the Lorentz force equation

$$\frac{d}{dt}(m\vec{V}) = q\left[\vec{E} + \vec{V} \times \vec{B}\right].$$

When the electrons hit the cavity wall, they will be terminated in ACE3P simulation and their phase space information will be written to a file, which will be used for postprocessing and further study.

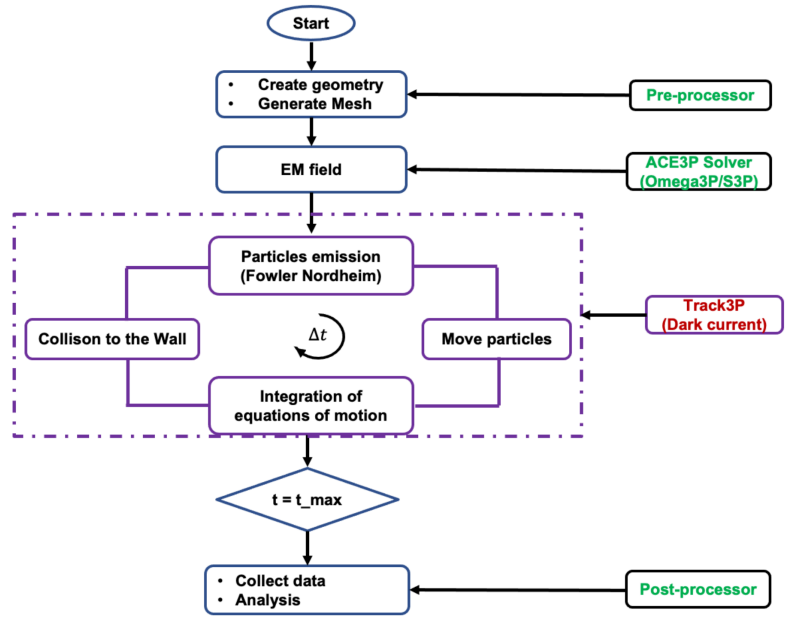

Figure 7. ACE3P dark current simulation workflow.

For a typical dark current simulation, a total of 80k primary particles are emitted from the surface and it takes 25 RF cycles for particles transiting the whole 56 cells. The computational time was less than 30 minutes to complete the end-to-end simulation using 8 CPU Perlmutter nodes for a 3.4M tetrahedral elements mesh. Fig. 8 shows a snapshot of particle trajectories. Fig. 9 presents a preliminary benchmark comparison of dark current between ACE3P simulations and measured data from KEK, showing a good overall agreement in shape. Further benchmarking will be conducted as additional measurement data become available.

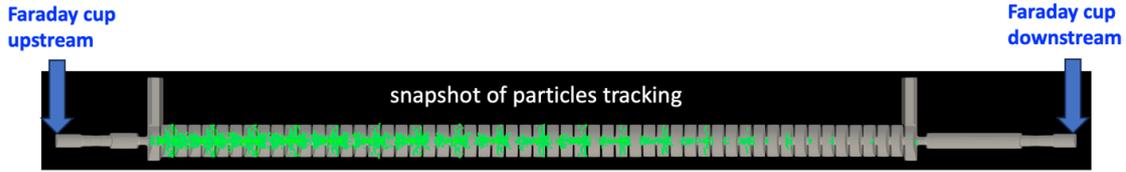

Figure 8. Snapshot of particle trajectories.

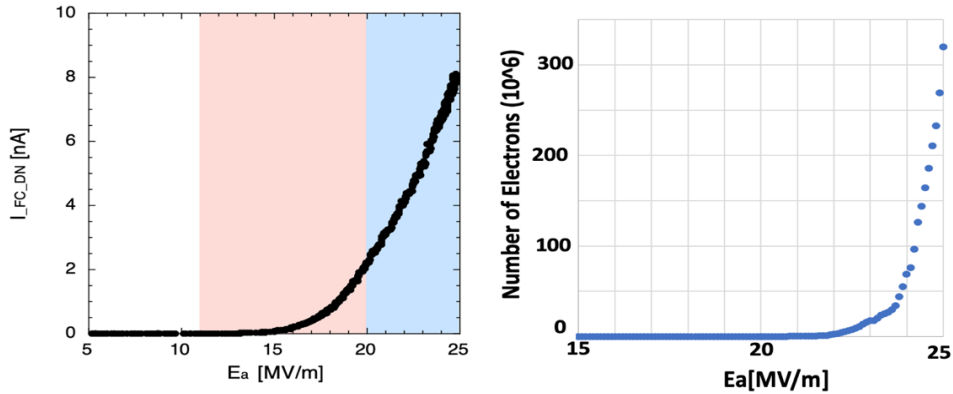

Figure 9. Dark current: Measured by KEK (Left); Calculated by Track3P (Right).

**3.3 Radiation study**

To conduct the radiation study using Geant4, the process begins with extracting the solid geometric model and its confined vacuum region from the KEK-provided design through Cubit. Fig. 10 shows the solid material model and its confined vacuum region. Track3P, a module in ACE3P, is used to simulate particle trajectories, with data collected for all particles impacting the cavity wall. Fig. 11 shows the initial particles distribution at 21.6 MV/m. These particles, along with the solid material model and vacuum region, are then loaded into Geant4. In this preliminary test, only particles with energy greater than 1 MeV (Fig.12) are loaded in Geant4. Finally, Geant4 is employed to simulate the interactions between the particles and the cavity wall, enabling a detailed study of the resulting radiation. Fig. 13 shows the radiation comparison between simulated by Geant4 and measured by KEK. While the overall curve aligns well, the scale discrepancy is attributed to an unknown measurement b, field enhancement factor. Additionally, the radiation contour plots in the bottom section of Fig. 13 highlight the regions of high radiation density, particularly in the downstream area.

**4. CONCLUSION AND FUTURE WORK**

An integrated simulation tool for studying dark current radiation effects has been developed, combining the capabilities of ACE3P and Geant4. This tool has been utilized to perform dark current and radiation studies, with a preliminary benchmark conducted between simulation results and measured data. The overall curve shows good agreement, with discrepancies in scale are attributed to unknown measurement parameters that differ from those used in the simulation. Further validation will be undertaken once additional measured data becomes available.

Future developments and applications of the integrated tool include:
- Developing a Python script to streamline the integrated simulation workflow
- Adding capabilities for positron source and capture simulations
- Integrating machine learning tools into the framework to enable inverse modeling using measured data

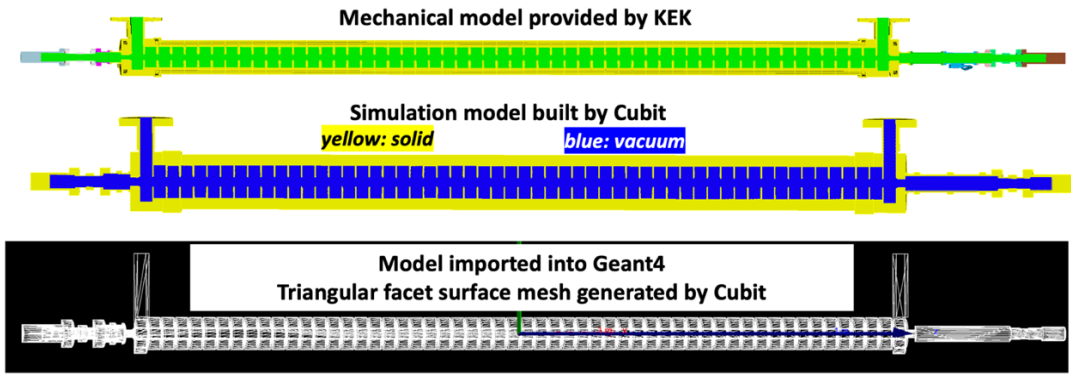

Figure 10. Geometric model load in Geant4 for radiation study.

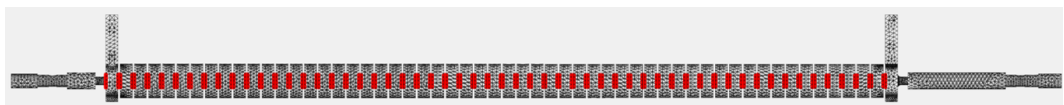

Figure 11. Initial particles distribution at 21.6 MV/m.

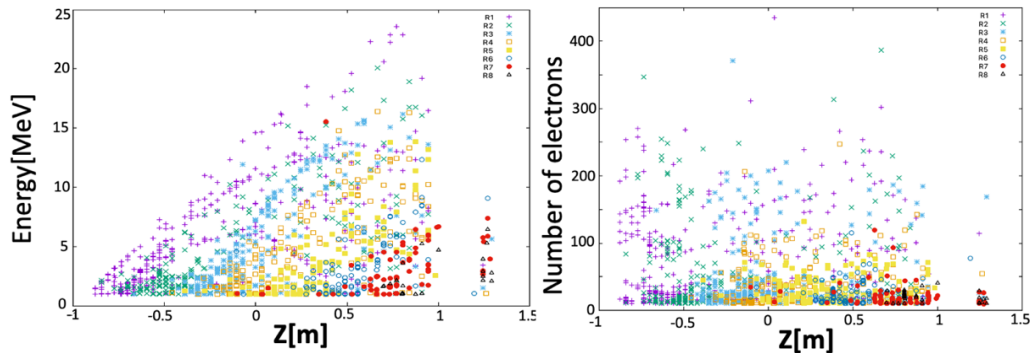

Figure 12. Particles with energy >1 MeV.

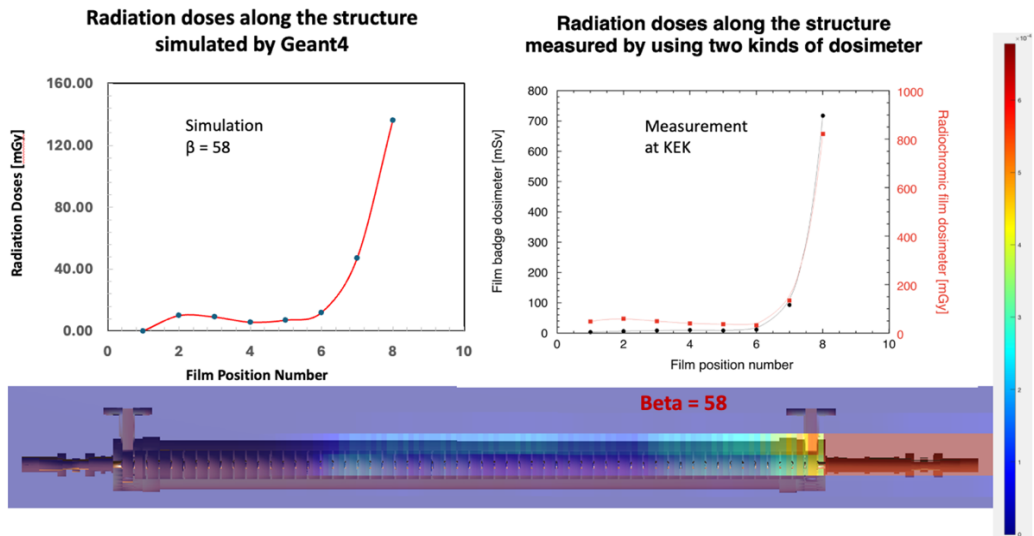

Figure 13. Radiation dose calculation by Geant4.

## Acknowledgments

This research used resources of the National Energy Research Scientific Computing (NERSC) Center, which is supported by the Office of Science of the U.S. Department of Energy under Contract No. DE-AC02-05CH11231.